\documentclass[12pt,letterpaper]{article}
\pdfoutput=1

\usepackage{amsmath,amssymb,calc, amsthm,bbm, epsfig,psfrag, mathtools}
\usepackage{graphicx}
\usepackage[numbers,sort&compress]{natbib}
\usepackage[dvipsnames]{xcolor}

\newcommand{\Comment}[1]{{}}
\definecolor{darkblue}{rgb}{0.15,0.35,0.55}
\definecolor{reddish}{rgb}{0.65, 0.2, 0.2}
\usepackage[linktocpage=true]{hyperref}
\hypersetup{
colorlinks=true,
citecolor=darkblue,
linkcolor=reddish,
urlcolor=darkblue,
pdfauthor={},
pdftitle={},
pdfsubject={}
}

\makeatletter
\renewcommand\section{\@startsection {section}{1}{\z@}%
                                   {-3.5ex \@plus -1ex \@minus -.2ex}
                                   {2.3ex \@plus.2ex}%
                                   {\normalfont\large\bfseries}}
\renewcommand\subsection{\@startsection{subsection}{2}{\z@}%
                                     {-3.25ex\@plus -1ex \@minus -.2ex}%
                                     {1.5ex \@plus .2ex}%
                                     {\normalfont\bfseries}}
\makeatother

\theoremstyle{plain}

\theoremstyle{definition}

\let\non\nonumber

\def\bea#1\eea{\begin{align}#1\end{align}}

\def\bes #1\ees{\begin{split}#1\end{split}}

\newcommand{\be}{\begin{equation}}
\newcommand{\ee}{\end{equation}}

\newcommand{\bma}{\begin{pmatrix}}
\newcommand{\ema}{\end{pmatrix}}

\newcommand{\half}{\frac{1}{2}}

\newcommand{\R}{{\mathbb R}}

\newcommand{\cR}{{\cal R}}

\newcommand{\wh}{\widehat}

 \newcommand{\cL}{{\mathcal{L}}}

\newcommand{\rd}{{\rm d}}
\def\mn{_{\mu \nu}}

\let\a=\alpha

\let\l=\lambda

\let\s=\sigma

\let\w=\wedge

\let\S=\Sigma

\def\th{\theta}

\def\l{\lambda}
\def\m{\mu}
\def\n{\nu}
\def\o{\omega}

\def\s{\sigma}

\def\M{{\mathcal M}}

\newcommand{\La}{\Lambda}

\newcommand{\dd}{\delta}
\newcommand{\com}[2]{{ \left[ #1, #2 \right] }}
\newcommand{\acom}[2]{{ \left\{ #1, #2 \right\} }}

\newcommand{\p}{\partial}

\def\com#1#2{{ \left[ #1, #2 \right] }}
\def\acom#1#2{{ \left\{ #1, #2 \right\} }}

\newcommand{\C}[1]{$(\ref{#1})$}

\def\IZ{\relax\ifmmode\mathchoice
{\hbox{\cmss Z\kern-.4em Z}}{\hbox{\cmss Z\kern-.4em Z}}
{\lower.9pt\hbox{\cmsss Z\kern-.4em Z}} {\lower1.2pt\hbox{\cmsss
Z\kern-.4em Z}}\else{\cmss Z\kern-.4em Z}\fi}
\def\IR{\relax{\rm I\kern-.18em R}}

\def\one{{\hbox{ 1\kern-.8mm l}}}

\newlength{\bredde}
\def\slash#1{\settowidth{\bredde}{$#1$}\ifmmode\,\raisebox{.15ex}{/}
\hspace*{-\bredde} #1\else$\,\raisebox{.15ex}{/}\hspace*{-\bredde}
#1$\fi}

\newsavebox{\zzzbar}
\sbox{\zzzbar}
  {\setlength{\unitlength}{0.9em}
  \begin{picture}(0.6,0.7)
  \thinlines
  \put(0,0){\line(1,0){0.6}}
  \put(0,0.75){\line(1,0){0.575}}
  \multiput(0,0)(0.0125,0.025){30}{\rule{0.3pt}{0.3pt}}
  \multiput(0.2,0)(0.0125,0.025){30}{\rule{0.3pt}{0.3pt}}
  \put(0,0.75){\line(0,-1){0.15}}
  \put(0.015,0.75){\line(0,-1){0.1}}
  \put(0.03,0.75){\line(0,-1){0.075}}
  \put(0.045,0.75){\line(0,-1){0.05}}
  \put(0.05,0.75){\line(0,-1){0.025}}
  \put(0.6,0){\line(0,1){0.15}}
  \put(0.585,0){\line(0,1){0.1}}
  \put(0.57,0){\line(0,1){0.075}}
  \put(0.555,0){\line(0,1){0.05}}
  \put(0.55,0){\line(0,1){0.025}}
  \end{picture}}

\newfont{\goth}{ygoth.tfm scaled 1200}                   

 \numberwithin{equation}{section}

\def\1{{(1)}}
\def\2{{(2)}}
\def\3{{(3)}}

\renewcommand{\wh}{\widehat}

\begin{document}
\begin{titlepage}

\begin{center}

{November 30, 2016}
\hfill         \phantom{xxx}  EFI-16-20

\vskip 2 cm {\Large \bf DBI from Gravity} 
\vskip 1.25 cm {Travis Maxfield and Savdeep Sethi}\non\\
{\vskip 0.5cm  {\it Enrico Fermi Institute, University of Chicago, Chicago, IL 60637, USA}}

\vskip 0.2 cm
{ Email:} \href{mailto:maxfield@uchicago.edu}{maxfield@uchicago.edu}, \href{mailto:sethi@uchicago.edu}{sethi@uchicago.edu}

\end{center}
\vskip 1 cm

\begin{abstract}
\baselineskip=18pt

We study the dynamics of gravitational lumps. By a lump, we mean a metric configuration that asymptotes to  a flat 
space-time. Such lumps emerge in string theory as strong coupling descriptions of D-branes. We provide a physical argument that the broken global symmetries of such a background, generated by certain large diffeomorphisms, constrain the dynamics of localized modes. These modes include the translation zero modes and any localized tensor modes. The constraints we find are gravitational analogues of those found in brane physics. For the example of a Taub-NUT metric in eleven-dimensional supergravity, we argue that a critical value for the electric field arises from standard gravity without higher derivative interactions.

\end{abstract}

\end{titlepage}


\section{Introduction} \label{intro}

The Dirac-Born-Infeld action is a beautiful closed form collection of higher derivative interactions describing the dynamics of a brane with a world-volume gauge-field.  Imagine a $p$-brane with world-volume $\S_{p+1}$ embedded in a $D+1$-space-time dimensional manifold $\M$ with metric $G_{MN}$. The embedding of the brane is described by maps $X: \S_{p+1} \rightarrow \M$. In local coordinates $X = X(\s^\m)$ where $\s^\m$ are local coordinates on the brane. The action takes the form,
\be\label{DBI}
S_{DBI} = - T_p \int d^{p+1}\s \, f(\phi) \sqrt{- \det{\left(g_{\m\n} + \alpha F_{\m\n}\right)}}.
\ee
There are two dimensionful parameters $T_p$ and $\alpha$, as well as a possible function of scalar fields $f(\phi)$.  The field strength $F$ is abelian, while the pulled-back metric takes the explicit form:
\be
g_{\m\n} = G_{MN} \p_\m X^M \p_\n X^N, \qquad M,N=0,\ldots, D. 
\ee
Specific values for the scales $(T_p, \alpha)$ and the function $f(\phi)$ arise in different settings. For example when studying D-branes in perturbative string theory, $\alpha = 2\pi\alpha'$,  where $\alpha'$ determines the tension of the fundamental string. The other parameters $T_p$ and $f(\phi)$ are also determined in terms of $\alpha'$ and the string dilaton field.  
We are going to be less concerned with the theory-dependent values of these parameters and more concerned with the extent to which symmetry dictates the DBI form~\C{DBI}. 

Imagine a flat, infinitely extended brane embedded in ambient flat Minkowski space-time. By definition, any such brane spontaneously breaks Poincar\'e symmetry:
\be\label{poincare}
ISO(D,1) \quad\rightarrow\quad ISO(p, 1). 
\ee
The breaking of translational symmetry guarantees the existence of $D-p$ universal scalar fields on the brane world-volume, which serve as Nambu-Goldstone (NG) bosons for the broken translations. These scalar modes arise from the maps $X(\s^\m)$ after suitable gauge-fixing. We will be concerned only with these universal scalars along with the gauge-field supported on the brane, and how these modes arise and interact in a purely gravitational setting.  

The reason for our interest can be motivated from string theory by considering a D6-brane moving in $D=10$ flat Minkowski space-time. This brane appears in the type IIA string. It has many nice properties, like preserving one-half of the ambient $32$ supersymmetries and spontaneously breaking $D=10$ Poincar\'e symmetry:
\be\label{poincare}
ISO(9,1) \quad\rightarrow\quad ISO(6, 1). 
\ee
However, the most important property for us is the appearance of the DBI action as the world-volume action on the brane at weak string coupling. More precisely, the Born-Infeld action can be derived by studying the conditions for world-sheet conformal invariance for a D-brane with a constant background electromagnetic field strength in flat Minkowski space-time~\cite{Abouelsaood:1986gd}.\footnote{In our subsequent discussion, we will sometimes loosely use DBI to refer to~\C{DBI}\ even with the gauge-fields set to zero. In that situation, the DBI action really reduces to the Dirac action for the universal scalar fields. On the other hand, the Born-Infeld action just refers to the non-linear action governing the gauge-fields with a flat Minkowski brane metric.} 

The Dirac action on this brane can also be derived from space-time symmetry arguments, as we will describe shortly. This is the key point for us. At strong coupling, the D6-brane has an exact realization as a Kaluza-Klein monopole, which is an example of a gravitational ``lump." The Kaluza-Klein monopole solution is constructed as follows: consider $D=11$ Minkowski space-time of the form $\R^{9,1}\times S^1$. In the context of string theory, we want to consider $D=11$ supergravity, but tensor fields beyond the metric do not play a role in the construction of the monopole background. Therefore, consider standard Einstein gravity on this space-time. From the metric reduced on $S^1$, we find a $D=10$ abelian gauge-field. Choose an $\R^3$ from the spatial directions of $\R^{9}$ and fiber the gauge-field over this $\R^3$ so that it has non-zero magnetic charge. This is a Kaluza-Klein monopole. 

The $4$-dimensional metric on $S^1\times \R^3$ is the Taub-NUT metric~\cite{Hawking:1976jb, Gibbons:1979xm}. It is smooth, although it has a coordinate singularity at the point where the circle shrinks to zero size; this point is the location of the brane. 
The full $11$-dimensional metric $G$ takes the form, 
\be\label{gravsoliton}
ds^2_{11} = ds^2_{6+1}(x) + ds^2_{TN}(y, \chi), 
\ee
where $ds^2_{6+1}$ is the flat $7$-dimensional Minkowski metric with coordinates $x^\m$ and, 
\be
ds^2_{TN} = H(y) d\vec{y}^2 + H(y)^{-1} \left( d\chi + \vec{A}\cdot d\vec{y} \right)^2, \qquad H(y) = 1 + {\ell\over |\vec{y}-\vec{y}_0|}. 
\ee
The $\vec{y}$ are coordinates for $\R^3$ and $\chi$ is the coordinate for $S^1$ with periodicity 
\be\chi \sim \chi + 4\pi\ell.\ee 
The metric depends on one function $H$ harmonic in $\R^3$ and one scale $\ell$. The location of this gravitational ``brane'' is determined by $\vec{y_0}$. The vector potential $\vec{A}$ is determined by the harmonic function $H$:
\be
\nabla H = \nabla \times \vec{A}. 
\ee
The metric is smooth, hyperK\"ahler and easily generalizable to many centers. While it appears to be a metric on $\R^3\times S^1$, the space is actually topologically $\R^4$ because the circle unwinds at the location of the brane. 

This metric is asymptotically locally flat (ALF); at spatial infinity, it asymptotes to a flat metric. This is a rather important point for us. In this sense, the metric~\C{gravsoliton}\ spontaneously breaks the same symmetries~\C{poincare}\ as the D6-brane, where $ISO(9,1)$ is part of the symmetry group of $\R^{9,1}\times S^1$. While this gravitational background is a solution of M-theory, it is also a solution of any string theory compactified on a circle, or for that matter, any standard theory of gravity compactified on a circle. This will be the main example we study in this work, though the arguments we present are not tied to string theory, or this particular metric. They are based on symmetries and should apply to any gravitational lump with a metric that asymptotes to flat Minkowski space-time. While we are restricting to lumps in Minkowski space-times, these symmetry arguments should generalize to lumps in any asymptotic space-time with some isometries, including cases that cannot be realized in perturbative string theory like de Sitter space. 

Instead of spelling out the match of degrees of freedom in general, let us just focus on the $3$ universal scalars of the D6-brane. Those scalars parametrize the position of the brane in the transverse $\R^3$. In the dual gravitational description, Taub-NUT has a modulus $\vec{y_0}$, which parametrizes its center. Associated to $\vec{y_0}$ are $3$ metric moduli. 
These zero mode fluctuations of the metric match the $3$ scalars on the D6-brane. At weak coupling, we expect the Dirac action to govern the scalars. What can we expect at strong coupling? In the absence of symmetry, there is little hope of finding an interesting answer. However, given that DBI is determined by spontaneously broken symmetries, we might expect that those same symmetries provide similar constraints for gravitational lumps. 

At first sight, this might seem unlikely. The standard way to determine the collective coordinate dynamics of a metric like~\C{gravsoliton}\  proceeds by promoting each collective mode $\vec{y_0}$ to a slowly-varying field $\vec{y_0}(x)$. Substituting into the $D=11$ Einstein Hilbert action, 
\be\label{EH}
S_{D=11} ={1\over 2\kappa_{11}^2} \int \sqrt{-G} R, 
\ee
and integrating over the Taub-NUT space gives a $2$ derivative effective action. For the Taub-NUT solution, past studies of the collective coordinate effective action include~\cite{Ruback:1986ag, Imamura:1997ss, Bergshoeff:1997gy}. There has also been a suggestion that the higher derivative terms found in the BI action might arise from higher curvature interactions in gravity~\cite{Gibbons:2000xe}. Indeed there are higher derivative gravitational interactions in M-theory and string theory beyond the $2$ derivative terms captured by the Einstein-Hilbert action. Yet if broken symmetries are the fundamental reason for the emergence of DBI on branes, we might hope that broken symmetries are also sufficient to give DBI for gravitational lumps just from the Einstein-Hilbert action~\C{EH}. 

We will provide a physical argument that this is indeed the case for the universal scalars, and we will provide evidence that this is also the case for the gauge-field. There is good reason to expect that two derivative gravity alone is sufficient to see the Born-Infeld action for the gauge-field. Maximal supersymmetry pairs the gauge-field with the scalars, and essentially implies the full DBI action if the scalars are governed by the Dirac action. We will not be using supersymmetry in this analysis. Rather, we are interested in the implications of the universal symmetries present in any gravitational theory.   

 In section~\ref{DBIforbranes}, we review how the Dirac action emerges for branes from broken symmetries. We also discuss the extent to which symmetries determine the dynamics of the brane gauge-field. For the gauge-field, we do not know of any space-time symmetry argument that implies the Born-Infeld action, but there are interesting constraints that have been obtained recently in~\cite{Gliozzi:2011hj, Casalbuoni:2011fq}.\footnote{If one is willing to use symmetries of string theory, like T-duality, then it is easy to derive the Born-Infeld action on D-branes. However, we are really interested in the universal space-time symmetries of a gravitational theory that do not depend on any specific ultraviolet completion. } These constraints require that Lorentz invariant operators involving the gauge field strength $F$ be constructed using the pulled-back metric $g_{\m\nu}$. However, this does not imply the Born-Infeld action;  for example, any power $F^{n}$ is in principle possible. 

In section~\ref{gravlumps}, we turn to gravitational lumps. We describe the relation between diffeomorphisms and global symmetries for general lumps. We then focus on our main example of a single centered Taub-NUT space. In this gravitational setting, we find an analogue of the inverse Higgs mechanism, which relates the Nambu-Goldstone (NG) bosons generated by different broken Poincar\'e generators~\cite{Ogievetsky:1974}. Unlike the brane setting, the relation involves the appearance of bulk modes. Using this constraint, we determine the action of the broken and unbroken global symmetries on the NG bosons. The relations match those found in the brane analysis, and imply that the Dirac equation governs these universal scalars. We also find explicit expressions for the zero mode metric fluctuations corresponding to these scalars. We then extend our analysis to include an abelian gauge-field. Once again, the action of the nonlinearly realized symmetries generates both localized and bulk modes. Our final result for the action on the localized gauge-field agrees with the brane analysis, which again is not sufficient to imply the Born-Infeld action. 


In section~\ref{critelectric}, we turn to the question of whether gravity can teach us more about the gauge-field equation of motion. A sharp distinction between a free Maxwell theory and the Born-Infeld theory is the existence of a critical electric field in the latter theory. Given the prior transformation properties of the gauge-field under the broken Poincar\'e symmetries, it is essentially clear that the Born-Infeld action will govern the gauge-field if gravity can see a critical electric field. To demonstrate that this is indeed the case, we construct the fully back-reacted gravity solution for a constant electric field. We use string theory dualities as a means of generating this solution. Indeed, a critical value for the electric field does emerge from the structure of the gravitational solution. This actually gives us more information than broken symmetries alone imply for the action on a brane.   

There is one other significant motivation for us to understand whether and how DBI emerges from Einstein-Hilbert gravity. The motivation follows from the field theories supported on branes with constant backgrounds that break Lorentz invariance. 
For example, a D6-brane with a background $B_2$-field in string theory can give rise to a non-commutative field theory supported on the brane. 
At strong coupling, this brane configuration becomes an interesting generalization of a Taub-NUT solution, closely related to the solution described in section~\ref{critelectric}. Can a non-commutative gauge theory emerge from this gravitational background with no non-abelian isometries? There are tantalizing hints that this might well be the case, found in a perturbative expansion~\cite{Dasgupta:2003us}. The sticking point in that past analysis was precisely this question of whether DBI could really emerge from a $2$ derivative theory like~\C{EH}. Given the results found here, we look forward to revisiting those questions with the hope of eventually understanding more about the mysterious $D=6$ $(2,0)$ interacting chiral tensor theory, which can be constructed as a gravitational lump in type IIB string theory.   

In closing, we should also mention that our analysis, which is purely gravitational, should have an interesting relation to the blackfolds program described in, for example,~\cite{Emparan:2009at, Niarchos:2014maa}. That program is more in the spirit of including leading order gravitational back-reaction around a brane configuration.  The consistency constraints of supergravity have also been shown to reproduce the full non-linear equations of DBI in interesting work heading toward a DBI/supergravity correspondence in~\cite{Lunin:2007mj, Lunin:2008tf}. The appearance of DBI from supergravity was also discussed in~\cite{Hatefi:2012bp}, motivated by~\cite{Park:2001bm}.

\section{Deriving DBI for Branes}   
\label{DBIforbranes}
The Dirac action for a brane can be derived from the symmetries of a hypersurface embedded in an ambient space-time. For simplicity we will start with the co-dimension 1 case, which was described in~\cite{deRham:2010eu,Goon:2011qf}, but the construction can be extended to higher co-dimension straightforwardly~\cite{Hinterbichler:2010xn,Dubovsky:2012sh,Aharony:2013ipa}. 

There is a closely related approach using the coset construction for nonlinearly realized symmetries. The standard coset formalism for global symmetries was extended to space-time symmetries in~\cite{volkov,Ogievetsky:1974}. 
The derivation of the Dirac action in this language has been worked out in a variety of places; for example, an explicit derivation can be found in~\cite{Gomis:2006xw}.  

\subsection{Determining the scalar action}\label{branescalars}

Imagine, for simplicity, a 3-brane embedded in a $(4+1)$-dimensional ambient space-time, ${\cal M}$, with coordinates $X^A$ and metric $G_{AB}$. The location of the brane in the higher-dimensional space-time is given by the embedding functions,
\be
X^A(\s): {\R}^{3,1} \hookrightarrow {\cal M}~,
\ee
where $\s^{\mu}$ are coordinates on the brane. The $X^A$ are the dynamical variables.  Tangent vectors to the brane have components $e^{A}_{\mu}=\frac{\partial X^{A}}{\partial \s^{\mu}}$. We can pull-back the metric in the ambient space to find the induced metric on the brane:
\begin{align}
g_{\mu\nu}&=e^{A}_{\mu}e^{B}_{\nu}G_{AB}\ .
\label{inducedmetricdef}
\end{align}
There is also a single normal vector to the brane, which is transverse to the tangent vectors and normalized:
\be
n^{A}e^{B}_{\mu}G_{AB}=0~,~~~~~~~~~~~~~~~n^{A}n^{B}G_{AB}=1.\
\ee
The normal and tangent vectors are used to construct the extrinsic curvature tensor,
\begin{align}
K_{\mu\nu}&=e^{A}_{\mu}e^{B}_{\nu}\nabla_{A}n_{B}\ ,
\end{align}
where $\nabla_A$ is the bulk covariant derivative. Brane diffeomorphisms act as follows,
\be
\delta^{\rm brane}_\xi X^A = \xi^\mu\partial_\mu X^A,
\ee
and we demand that the world-volume action be invariant under these reparameterizations. This requires the action to be a diffeomorphism scalar constructed from the geometric ingredients
\be
S=\int {\rm d}^{4}x\, \sqrt{-g}\,\mathcal{L}(g_{\mu\nu},\nabla_\mu, R^\mu_{~\nu\rho\sigma}, K_{\mu\nu} )\ .\label{genericaction}
\ee
Note that $R_{\mu\nu\rho\sigma}$ and $K_{\mu\nu}$ are not independent ingredients because of the Gauss--Codazzi relation,
\be
R_{\mu\nu\rho\sigma} = K_{\mu\rho}K_{\nu\sigma} - K_{\nu\rho}K_{\mu\sigma}, 
\ee
which determines $R$ in terms of $K$. It is therefore convenient to take $K_{\mu\nu}$ as the fundamental building block.

We now want to fix the gauge symmetries of this action. In order to accomplish this, we choose a foliation of the bulk by surfaces of constant $X^4$ and adapt the worldvolume coordinates to this foliation by choosing static (Monge) gauge
\begin{align}
X^{\mu}(\s)=\s^{\mu}, \quad \m=0,\ldots, 3\qquad X^4(\s)=\phi(\s),\ \ \label{gaugechoice}
\end{align}
This is the usual gauge choice that identifies the $4$ world-volume coordinates with bulk coordinates. The remaining coordinate, $\phi$, represents the dynamical degree of freedom of the brane.

If the ambient space-time has any isometries, the brane action will inherit global symmetries. For each bulk Killing vector $K$ satisfying,
\be
\cL_K G_{AB}  = K^C\partial_C G_{AB}+\partial_A K^C G_{CB}+\partial_BK^CG_{AC} = 0,
\ee
the action inherits a global symmetry:
\be
\delta^{\rm bulk}_K X^A = K^A.
\label{globalsymm}
\ee
The pull-back of the bulk metric and the extrinsic curvature are both invariant under this transformation, which implies invariance of the action~\eqref{genericaction}. However, the transformation~\eqref{globalsymm}\ generically does not preserve the gauge choice~\eqref{gaugechoice}. Rather,
\be
X^\mu = \s^{\mu}\mapsto \s^{\mu}+K^{\mu}.
\ee
To maintain our preferred gauge~\C{gaugechoice}, we must perform a compensating brane diffeomorphism
\be
\delta^{\rm brane}_{-K} X^A = -K^\mu\partial_\mu X^A,
\ee
so that the combined transformation $\dd_K\equiv (\delta^{\rm bulk}_K+\delta^{\rm brane}_{-K})$,
\be
\dd_K X^\mu = 0, \qquad \dd_K \phi = -K^\mu\partial_\mu\phi + K^4,
\label{gaugefixedglobalsymms}
\ee
is a global symmetry of the gauge-fixed action. 

To derive the Dirac action in its traditional form, we choose a flat ambient space-time ${\cal M} = {\mathbb R}^{4,1}$. A convenient choice of coordinates for the bulk is given by,
\be
ds^2 = \left(dX^4\right)^2 + \eta_{\mu\nu}dX^\mu dX^\nu,
\ee
where the bulk is foliated by constant $X^4$ slices. The induced metric and extrinsic curvature in static gauge are given by~\cite{Goon:2011qf}
\begin{align}
g_{\mu\nu} = \eta_{\mu\nu}+\partial_\mu\phi\partial_\nu\phi,\qquad K_{\mu\nu} = -\frac{\partial_\mu\partial_\nu\phi}{\sqrt{1+(\partial\phi)^2}}. 
\label{eq:inducedgandk}
\end{align}
From the perspective of a reparametrization invariant action, the lowest order action in a derivative expansion is simply, 
\be
S = \int{\rm d}^4\s \sqrt{-g}.
\ee
On the other hand, after gauge-fixing this action takes the form: 
\be
S = \int{\rm d}^4\s \sqrt{-g} = S = \int{\rm d}^4\s\sqrt{1+(\partial\phi)^2}.
\ee
This is the lowest order action that enjoys the full $ISO(4,1)$ symmetry. Following~\cite{Aharony:2013ipa}, we define the weight of an operator as the number of derivatives minus the number of scalar fields. We might wonder whether the constraints imposed by broken Poincar\'e invariance are equivalent to the constraints imposed by reparametrization invariance for branes in flat Minkowski space-time. Indeed, an argument for this equivalence for the universal scalars appears in~\cite{Cooper:2013kga}. 

In order to see the global symmetries of this action, we note that the $5$-dimensional ambient Minkowski space-time has $15$ Killing vectors,
\begin{align}
P_A = -\partial_A , \qquad J^{AB} = X^A\partial^B-X^B\partial^A,
\end{align}
corresponding to $5$ translations and $10$ rotations, which act on the coordinates as follows:
\be
\delta^{\rm bulk}_{P_B}X^A = \delta^A{}_B, \qquad \delta^{\rm bulk}_{J^{BC}} X^A = X^C \delta^{AB} - X^B \delta^{AC}
\ee
Under a general Poincar\'e transformation generated by $c^AP_A + \Lambda_{AB} J^{AB}$, we note that
\be
\delta^{\rm bulk} X^\mu = \Lambda^\mu_{~\nu}X^\nu + \Lambda^\mu_{~4}X^4 +c^\mu.
\ee
Under the action of $P_4$, we see that $X^\mu$ does not transform. However, it does transform under $J^{\mu4}$ so we require a compensating brane diffeomorphism to maintain static gauge. 
\be
\delta^{\rm brane} X^\mu  = -\Lambda^\mu_{~4} X^4. 
\ee
Using~\eqref{gaugefixedglobalsymms}, we see that the global symmetries of the gauge-fixed action are
\begin{align} \label{branesymmetries}
\dd_{P_4}\phi = c^4, \qquad \dd_{J_{\mu 4}}\phi = \s^\mu + \phi\partial^\mu\phi,
\end{align}
with the 10 symmetries, $P_\mu = -\partial_\mu$, $J^{\mu\nu} = \s^\mu\partial^\nu-\s^\nu\partial^\mu$, realized linearly on $\phi$. The first symmetry of~\C{branesymmetries}, broken translation invariance, simply requires each $\phi$ to appear with a derivative and thus forbids any potential for $\phi$. The really powerful transformation is the second symmetry of~\C{branesymmetries}, broken Lorentz invariance, which forbids a free scalar action
$$
S = \int (\p \phi)^2, 
$$
and requires the higher derivative interactions that appear in the Dirac action~\C{DBI}. 
%
%
%
%
%
%

For later use we note that the induced metric,
\be\non
g_{\mu\nu} = \eta_{\mu\nu}+\partial_\mu\phi\partial_\nu\phi,
\ee
has the following transformation under a broken Lorentz symmetry: imagine boosting the transverse direction into a world-volume direction specified by a vector field $v_\m$. Under this broken symmetry,  we see that 
\be\label{metrictransform}
\dd_{v^\m J_{\mu 4}} g_{\mu\nu} = \left(\phi v^\l\right)\partial_\l g_{\mu\nu} + g_{\l\nu}\partial_\mu\left(\phi v^\l\right)+g_{\l\mu}\partial_\nu\left(\phi v^\l\right).
\ee
We recognize this transformation as the action of  the Lie derivative, $\cL_{\phi v^\mu}$, along the direction specified by the vector $\phi v^\mu$.

\subsection{Including the gauge field} \label{includinggaugefields}
Given that the lowest weight action for the scalars is nicely determined by broken Poincar\'e invariance, it is natural to ask whether the full DBI action~\C{DBI}\ including the gauge-field might also follow from space-time symmetries. 

There has been interesting progress in this direction initiated in~\cite{Gliozzi:2011hj}, and nicely clarified in~\cite{Casalbuoni:2011fq}. In addition to the Nambu-Goldstone modes, imagine an abelian gauge-field, $A_\m$, supported on the brane world-volume. This is a quite different perspective on the world-volume gauge field than the usual string theory perspective, where both the universal scalars and the world-volume gauge field arise from the dimensional reduction of a $D=10$ vector field. 

The key observation is that this brane gauge field must also transform in order to nonlinearly realize the higher-dimensional Poincar\'e symmetry. Under $D=4$ translations and Lorentz rotations, $A_\mu$ transforms as a $1$-form. Under translations and boosts into the fourth direction, $A_\mu$ does {\it not} transform:
\be
\delta_{P_4}A =0, \qquad \delta_{J_{\mu4}}A =0.
\ee
However, as described above, $J_{\mu4}$ must be accompanied by a compensating world-volume diffeomorphism under which $A_\mu$ transforms as a $1$-form:
\be\label{branegaugetransformation}
\delta A_\mu = -\Lambda^\l_{~4}\left(\phi\,\partial_\l A_\mu +A_\l\partial_\mu\phi\right).
\ee
Note that under this symmetry the field strength transforms, 
\be
\dd_{v^\m J_{\mu 4}} F_{\mu\nu} = \left(\phi v^\l\right)\partial_\l F_{\mu\nu} + F_{\l\nu}\partial_\mu\left(\phi v^\l\right)+ F_{\mu\l}\partial_\nu\left(\phi v^\l\right),
\ee
again by the action of the Lie derivative along the vector $\phi v^\mu$, just like the brane metric~\C{metrictransform}. We therefore see that any diffeomorphism invariant object constructed from $g_{\mu\nu}$ and $F_{\mu\nu}$ will nonlinearly realize the $D=5$ Poincar\'e symmetries.

It is true that the linear combination $g_{\mu\nu} + \lambda F_{\mu\nu}$, with $\l$ an arbitrary parameter, transforms like a tensor. Diffeomorphism invariants constructed from this object will be invariant under the ambient space Poincar\'e symmetries. A natural low order invariant is the combination, 
\be
S= \int{\rm d}^4x \sqrt{{\rm det}\left(g_{\mu\nu}+\lambda F_{\mu\nu}\right)},
\ee
which is just the normal DBI action. Unfortunately, this is not a unique choice. For example, the combination $F_{\m\n} F^{\m\n}$ can also be used to build a good coupling if the indices are raised with the full pulled-back metric. 

A way to understand this goes as follows: both $g$ and $F$ transform the same way under the broken Poincar\'e generators. If one only has $g$ to build diffeomorphism invariants then the lowest weight action is uniquely the Dirac action. With both $g$ and $F$, any diffeomorphism invariant combination will automatically respect the broken Poincar\'e symmetries. To uniquely select the DBI action, one needs something more: either a larger symmetry like supersymmetry or some additional input. 

Despite not uniquely selecting the DBI action, the nonlinearly realized symmetries do require interesting interactions between any additional fields in the theory and the Nambu-Goldstone modes. The argument of~\cite{Gliozzi:2011hj}\ can be extended to an arbitrary field, either bulk or brane, transforming in a representation of either $ISO(D,1)$ or $ISO(p,1)$ for a $p$-brane in a $D+1$-dimensional space-time~\cite{Casalbuoni:2011fq}. We suspect this will give interesting data about the couplings on D-branes, only a subset of which are currently understood.

\section{Gravitational Lumps}\label{gravlumps}
\subsection{Diffeomorphisms and global symmetries}

We now turn to gravitational lumps. The picture is quite different from a brane because there is no analogue of brane world-volume coordinates, $\s^\m$, distinct from space-time coordinates, or a notion of an embedding. Rather there is simply a non-compact space-time metric on which we Kaluza-Klein reduce to find localized modes.    
Consider a $D+1$-dimensional space-time metric of product form, 
\be\label{fullmetric}
ds^2 = g_{MN} dX^M dX^N = \eta_{\mu \nu} dx^\mu dx^\nu + g_{ij}(y) dy^idy^j.
\ee
The product form of the metric could be generalized to a warped metric, but for simplicity, we will retain the simple product structure. The Minkowski space-time metric is $p+1$-dimensional. This is the analogue of the brane world-volume. The assumed product structure for~\C{fullmetric}\  describes a static lump. Boosted lumps will typically involve off-diagonal metric terms. 
We will allow circle identifications for the asymptotic metric so that solutions like Kaluza-Klein monopoles are permitted, but no more general compactifications. The Euclidean metric $g_{ij}$ is therefore assumed to be asymptotically flat:
\be\label{decay}
g_{ij}(y) \to \delta_{ij}  + \mathcal{O}(y^{-1}) 
\quad \text{ as } \quad |{y}| \to \infty.
\ee
We will not be very precise about necessary decay conditions on the metric. The decay rate~\C{decay}\ will be sufficient for our discussion, and germane to a large class of examples.  
 
The space-time possesses the asymptotic isometry group
\be
G = SO(d,1) \ltimes \mathbb{R}^{d,1} \times U(1)^{D-d} = ISO(d,1)\times U(1)^{D-d}. 
\ee
This group is reduced from $ISO(D,1)$ because of the circle identifications $U(1)^{D-d}$. Only a subgroup, $H$, acts as isometries away from infinity:
\be
H = ISO(p,1) \times \wh{G},
\ee
where $p+1$ is the dimension of the Minkowski space-time of~\C{fullmetric}\ and $\wh{G}$ is the isometry group of the metric $g_{ij}$. We will assume that $g_{ij}$ breaks all the non-compact translational symmetries, $\R^{d-p}$, that act on the $y$ coordinates at infinity.

In a gauge theory like gravity, gauge transformations are generated by compact diffeomorphisms which asymptote to the identity at infinity. These transformations identify points in the configuration space of fields. On the other hand, the asymptotic symmetry group, generated by large diffeomorphisms, acts like a global symmetry group---rotating distinct states into each other rather than identifying them. The subgroup $H$ then consists of those global symmetries that are linearly realized in the background, while the coset $G/H$ must be nonlinearly realized. By our assumptions, this coset includes the generators of translations in the $y$-coordinates, which we denote by $P_i$, as well as those generators that rotate $y$- and $x$-coordinates into each other. We denote these generators by $M_{\mu i}$ and call them, generically, boosts.

To the generators of broken symmetries, we expect associated Nambu-Goldstone bosons. For global symmetries, this association is one to one. However because these are space-time symmetry generators, the association between NG-bosons and broken generators is not one to one. Rather the NG-bosons corresponding to the $M_{\mu i}$ are not independent of those corresponding to the $P_i$ generators; this dependency is sometimes called the inverse Higgs constraint~\cite{Ivanov:1975zq}. In our gravitational setting, an analogue of the inverse Higgs mechanism arises in a somewhat different way. 
Let us explore precisely how in an explicit family of metrics.

\subsection{Gravitational instantons }\label{gravinstantons}

No condition that we have imposed so far on the metric~\C{fullmetric}\ would lead to a reasonable definition of a gravitational lump or brane. We would like the metric to admit normalizable zero modes that give rise to $p+1$-dimensional fields. Those localized modes are the analogue of brane world-volume fields. So let us focus on our primary example of the Taub-NUT metric which, as we discussed in section~\ref{intro}, is well motivated from string theory.    

The Taub-NUT metric falls into a larger class of complete, self-dual or anti-self-dual metrics on $\mathbb{R}^4$, known as Gibbons-Hawking spaces. The self or anti-self-duality of the curvature $2$-forms for these metrics is one reason that these Euclidean four-dimensional solutions can be viewed as gravitational instantons. They take the general form~\cite{Gibbons:1979xm}: 
\be
ds^2 = H(\vec{y}) d\vec{y}^2 + {1 \over H(\vec{y})} \left( d\chi^2 + \vec{A}(\vec{y}) \cdot d\vec{y} \right)^2,
\ee
where
\be
{\nabla} H = \pm {\nabla} \times \vec{A} \qquad \Rightarrow \quad { \nabla}^2 H = 0.
\ee
The $\pm$ correlates with self- and anti-self-duality, respectively. A general solution with $k$ centers is given by, 
\be
H(\vec{y} ) = \epsilon +  \sum\limits_{i=1}^k {\ell_{i-1} \over |\vec{y} - \vec{y}_{i-1}|}.
\ee
The apparent singularities at $\vec{y} = \vec{y}_i$ are removable `nut' singularities only if $\ell_{i-1} = \ell$ and the coordinate $\chi$ has periodicity $4\pi {\ell\over  k}$~\cite{Eguchi:1978gw}. The parameter $\epsilon \geq 0$ can be transformed to either $1$ or $0$ by a constant rescaling of the metric. So there are two families of solutions depending on the values of $(\epsilon, k)$:
\be
EH_k: \, \ (0,k), \qquad\qquad  TN_k: \,\ (1, k).
\ee
The first Eguchi-Hanson family of metrics, $EH_k$, 
is ALE, with asymptotic volume growth matching that of the flat metric on $\mathbb{R}^4$. 
The boundary of $EH_k$ is given by the lens space $L(k, 1)$. These spaces comprise the $A$-series of a general $ADE$ classification of ALE spaces.

The second family of multi-centered Taub-NUT spaces, $TN_k$, are ALF spaces with asymptotic volume growth like that of $\mathbb{R}^3$. 
The boundary of $TN_k$ is given by the total space of an $S^1$ fibration over $S^2$ with topological charge $k$. For $k = 0$, the boundary is $S^1 \times S^2$, while for $k = 1$ we find the Hopf fibration of $S^3$. For $k \geq 1$, the boundary is again the lens space $L(k,1)$ mentioned previously.


Let us focus our attention on the simplest example of a single-centered Taub-NUT metric. For the moment, we will just consider this $4$-dimensional metric without appending a Minkowski space-time. A perturbation $h$ of a background metric $g$ is a zero-mode (to first order) if it satisfies the linearized Einstein equation. For a transverse  traceless perturbation, this equation reads
\be
\Delta_L h_{ij} = \nabla^p \nabla_p h_{ij} - 2 R^p{}_{ij}{}^q h_{pq} = 0,
\ee
where $\nabla$ and $R$ are defined with respect to $g$, and $\Delta_L$ is the Lichnerowicz operator.  This condition follows  from the quadratic action for the fluctuation $h$,
\be\label{quadaction}
S[g + h] = \int d^4y \sqrt{|g|} ~R[g] + \int d^4y \sqrt{|g|}\left( \nabla^p h^{mn} \nabla_p h_{mn} + 2 R^p{}_{mn}{}^q h^{mn} h_{pq} \right) + \ldots.
\ee
It is easier to do calculations with the Taub-NUT metric written in spherical coordinates so let us introduce spherical coordinates $(y, \th, \psi, \chi)$, 
\be\label{refinedansatz}
ds^2_{TN} =H(y)  \left( dy^2 + y^2 d\th^2 + y^2 \sin\th^2 d\psi^2\right)  + H(y)^{-1}\left( d\chi + \cos\th d\psi\right)^2.
\ee
This is the metric with the nut singularity located at ${\vec y}_0 =0$. We have set $\ell=1$ so that,
\be
H(y) = 1 + {1\over y}. 
\ee
 We can choose coordinates so that the nut singularity is located at an arbitrary point ${\vec y}_0$. These three collective coordinates describe a family of metrics with the same action. Let us denote this family of metrics by $g_{ij}(y; {\vec y}_0)$ to spell out this dependence. 

Metric perturbations which correspond to derivatives with respect to collective coordinates are guaranteed to be zero modes; however, they are not guaranteed to be normalizable. As can be checked explicitly, the perturbations
\be
\delta g_{mn}^{(i)} = {\partial \over \partial y_0^i} \, g_{mn}(y; {\vec y}_0), \quad i=1,2,3,
\ee
are not $L^2$ normalizable with respect to the inner product, 
\be\label{innerprod}
\langle h_{mn} , h_{pq} \rangle := \int d^4 y \sqrt{|g|} ~h_{mn} h_{pq} g^{mp} g^{nq}.
\ee
acting on transverse, $\nabla^m h_{mn}=0$, and traceless perturbations~\cite{Gibbons:1978ji}. Instead, the $L^2$ normalizable zero-modes differ from the $\delta g^{(i)}$ above by compactly supported diffeomorphisms, which are pure gauge fluctuations. The normalizable zero-modes take the form
\be\label{momzeromode}
h^{(i)}_{mn} = {\partial \over \partial y_0^i} g_{mn}(y; {\vec y}_0) + \mathcal{L}_{\eta_i} g_{mn}, 
\ee
where $\eta_i$ is a compactly supported vector field.

A convenient way to derive the zero-modes is from a set of harmonic functions. This observation was used in~\cite{Gibbons:1978ji}\ to find zero modes for the Schwarzschild  metric. Suppose $f^{(i)}$ is a harmonic function on Taub-NUT and that the gradient of $f^{(i)}$ asymptotes to an asymptotic Killing vector
\be
\nabla^m f^{(i)} \to \left( {\partial \over \partial y^i }\right)^m.
\ee
In this case the metric perturbation,
\be
h^{(i)}_{mn} = \nabla_m \nabla_n f^{(i)},
\ee
is transverse and traceless:
\be
\nabla^m h^{(i)}_{mn} = g^{mn}h^{(i)}_{mn} = 0.
\ee
It is also guaranteed to be a zero mode because it is the combination of a true diffeomorphism along with an asymptotic isometry. 

A set of such functions are the coordinates themselves:
\be
\{ f^{(1)}, f^{(2)}, f^{(3)} \} = \{ y\sin\theta \cos\psi, y\sin\theta\sin\psi, y\cos\theta \} = \{y_1, y_2, y_3\}.
\ee
Somewhat surprisingly, we find that the coordinates themselves are harmonic on the full Taub-NUT metric:
\be
\nabla^2 f^{(i)} = 0.
\ee
It is worthwhile making explicit the vectors that generate the zero modes:
\be
\bes
g^{mn} \nabla_n f^{(1)} &= {1 \over H} \left({\partial \over \partial y_1}\right)^m - {1 \over H} {\cot\theta \sin\psi \over y} \left( {\partial \over \partial \chi} \right)^m, \cr
g^{mn} \nabla_n f^{(2)} &= {1 \over H} \left({\partial \over \partial y_2}\right)^m + {1 \over H} {\cot\theta \cos\psi \over y} \left( {\partial \over \partial \chi} \right)^m, \cr
g^{mn} \nabla_n f^{(3)} &={1 \over H} \left({\partial \over \partial y_3}\right)^m. 
\ees
\ee
These vectors asymptote to the Killing vectors and rotate as a triplet under the $SO(3)$ isometry. Furthermore, the perturbations derived from the $f^{(i)}$ are normalizable and have the following $L^2$ norm,
\be
\langle h_{mn}^{(i) }, h^{(j)}_{pq} \rangle = 8\pi^2 \delta^{i j}.
\ee

\subsection{Motion on the moduli space}

So far we have only described zero mode fluctuations of the Taub-NUT metric. We now enlarge the space-time to include $p+1$-directions orthogonal to the Taub-NUT space with a Minkowski metric:
\be
ds^2 = \eta_{\mu \nu} dx^\mu dx^\nu + ds^2_{TN} := g_{MN} dX^M dX^N.
\ee
The previously mentioned described zero-modes of Taub-NUT are still zero-modes, and they are still transverse and traceless. However, we would like to allow the collective coordinates of Taub-NUT to slowly vary over the rest of space-time. We do this by considering a new perturbation
\be
\delta g_{MN} = \begin{pmatrix} 0 & 0 \\ 0 & \sum_i \phi^i(x) h^{(i)}_{mn} \end{pmatrix},
\ee
where we have split the perturbation according to the index split $M \to (\mu, m)$. This is still traceless and transverse because the total space is a product metric and $\phi$ depends only on $x$. Therefore, it will remain a zero mode if
\be
\nabla^P \nabla_P \delta g_{MN} - 2 R^P{}_{MN}{}^Q \delta g_{PQ} = 0,
\ee
which implies \be \partial^\mu \partial_\mu \phi^i(x) = 0. \ee 
This equation of motion can also be derived by plugging the perturbation $\delta g_{MN}$ into a quadratic action similar to eq.~(\ref{quadaction}) and integrating over the Taub-NUT directions to arrive at an effective lower-dimensional action. The use of the normalizable $h^{(i)}$ is necessary for that procedure. We see Klein-Gordon emerge to leading order in metric fluctuations. This is typically as much as can be extracted from the moduli space approximation. To get higher velocity interactions, we will need to understand the symmetries of the theory in more detail.

\subsection{The inverse Higgs constraint}\label{inversehiggs}
We have already found, to first order, the metric fluctuations that generate the NG-bosons corresponding to broken translations. Up to a gauge transformation, they are given by,  
\be\label{metricfluct}
\delta g_{MN} = \phi^i(x) \mathcal{L}_{P_i} g_{MN}.
\ee
We saw in section~\ref{gravinstantons}\ that this mode is not normalizable unless we modify it by compactly supported diffeomorphisms, so we will view~\C{metricfluct}\ as in the same gauge equivalence class as the leading order normalizable zero-mode.

The perturbation~\C{metricfluct}\ changes the metric to
\be
ds^2 = g_{MN} dX^M dX^N = \eta_{\mu \nu} dx^\mu dx^\nu + \left( g_{ij}(y) + \phi^k(x) \mathcal{L}_{P_k} g_{ij}(y) \right) dy^idy^j,
\ee
This metric is still in a warped product form so that the coordinate vectors associated with $x^\mu$ and $y^i$ are orthogonal. This is how we distinguish `internal' from `space-time' directions. 

Just like the case of broken translations, the NG-bosons corresponding to boosts should be given, at lowest order, by
\be
\delta g_{MN} = K^{\mu i}(x) \mathcal{L}_{M_{\mu i}} g_{MN}.
\ee
This fluctuation does not, however, preserve the warped product structure. Define the vector $M_{\mu i}^a$ by
\be
M_{\mu i}^a =x_\mu \left( {\partial \over \partial y^i}\right)^a -  y_i \left( {\partial \over \partial x^\mu} \right)^a,
\ee
where we raise and lower the $\mu$- and $i$-indices on the coordinates with $\eta_{\mu \nu}$ and $\delta_{ij}$ respectively. The Lie derivative along this vector-field changes the metric as follows:
\be
\bes
\mathcal{L}_{M_{\mu i}} g_{\nu \l} &= 0, \cr
\mathcal{L}_{M_{\mu i}} g_{\nu j} &= \left(g_{ij} - \delta_{ij} \right) \eta_{\mu \nu}, \cr
\mathcal{L}_{M_{\mu i}} g_{jk} & = x^\mu \mathcal{L}_{P_i} g_{jk}. \label{Maction}
\ees
\ee
Note that each term on the right hand side vanishes as $y\rightarrow \infty$. Note also that apart from the cross-term, the action of $M_{\mu i}$ on the metric is identical to that of the translation perturbation with $\phi^i(x) = x^\mu$. So we need to understand the meaning of the second line of~\C{Maction}.  The right hand side of~\C{Maction}\ defines a metric perturbation with components $(h_{\nu j}, h_{jk})$.  The fluctuation $\mathcal{L}_{P_i} g_{jk}$ for large $y$ decays like ${1\over y^2}$. On the other hand, the asymptotic metric $g_{ij}$ tends to a flat metric as $|y|\rightarrow \infty$ so the fluctuation $h_{jk}$, which is proportional to $\mathcal{L}_{P_i} g_{jk}$ is normalizable in the asymptotic region with respect to the inner product~\C{innerprod}:
\be
\langle h_{jk} , h_{lm} \rangle := \int d^4 y \sqrt{|g|} ~h_{jk} h^{jk} \approx\int_{|y|\to\infty}  y^2 dy\,  {1\over y^4} <\infty. 
\ee
On the other hand, $h_{\n j} \sim \left(g_{ij} - \delta_{ij} \right)$ behaves like ${1\over y}$ for large $y$ so the fluctuation is not normalizable: 
\be
\langle h_{\n j} , h_{\rho k} \rangle := \int d^4 y \sqrt{|g|} ~h_{\n j} h^{\n j} \approx\int_{|y|\to\infty}  y^2 dy\,  {1\over y^2} \rightarrow \infty. 
\ee
More importantly, it cannot be made normalizable by adding any compactly supported diffeomorphism. 

This is the first appearance of a novelty found in the action of the spontaneously broken Poincar\'e generators in a gravitational theory: they act not only on the normalizable `brane' modes, but also on the non-normalizable bulk modes. We will see this phenomenon again shortly. In this setting, because of the asymptotic circle the bulk modes are effectively $p+4$-dimensional fields while localized modes are $p+1$-dimensional fields. 

To actually generate a normalizable zero mode, we should modify $M_{\mu i}$ by a compactly supported diffeomorphism similar to that appearing in~\C{momzeromode}. This is equivalent to modifying the asymptotically normalizable portion of the metric perturbation to be transverse and traceless, which amounts to:
\be
\mathcal{L}_{M_{\mu i}} g_{jk} = x^\mu \mathcal{L}_{P_{i}} g_{jk} \to x^\mu h^{(i)}_{jk}, 
\ee  
where $h^{(i)}_{jk}$ is one of the three normalizable perturbations appearing previously. We can now state the gravitational analogue of the inverse Higgs mechanism: namely, the action of $M_{\mu i}$ on the metric generates a normalizable zero mode proportional to the zero mode generated by $P_i$, but it also generates an accompanying non-normalizable bulk metric perturbation.


\subsubsection{Transformations under $H$ and $G/H$}
Now we have seen that, at lowest order, the normalizable metric perturbations describing the independent NG-bosons are entirely captured by those of translations; namely, 
\be
\delta g_{MN} = \phi^i(x) \mathcal{L}_{P_i} g_{MN},
\ee
up to a gauge transformation. We would next like to ask how the $\phi^i(x)$ transform under both the broken and and unbroken symmetries. To answer this question, we first need to generate an ``embedding'' for our gravitational lump. An embedding for a brane is described by the space-time variation of the universal scalars $\phi^i(x)$ that determine the location of the brane in the transverse space. So we will wiggle our metric by allowing the center of the Taub-NUT space to vary as a function of the space-time $x$-coordinates. This perturbed metric takes the form, 
\be\label{wiggle}
g_{MN}(\phi) = e^{\phi^i(x) \mathcal{L}_{P_i} } g_{MN}.
\ee
We do not expect this to be the unique way to wiggle the metric, but it is a physically well motivated perturbation. The metric~\C{wiggle}\ is not generally a solution of Einstein's equations, but determining the action of symmetries should not require an on-shell background. The metric does now depend on $\phi^i(x)$ and we can study the action of the generators of $G$ on the perturbed metric~\C{wiggle}\ in order to determine how those generators act on the $\phi^i$ fields. 

For example, a translation by a constant $\epsilon a^i$ for the $y^i$-coordinate acts as follows,
\be
\bes
e^{\epsilon a^i \mathcal{L}_{P_i}}e^{\phi^i(x) \mathcal{L}_{P_i} } g_{MN} &= e^{\left( \phi^i(x) + \epsilon a^i\right) \mathcal{L}_{P_i}} g_{MN}, \cr
&= g_{MN}(\phi + \epsilon a).
\ees
\ee
Now consider a boost with parameter  $\epsilon \Lambda^{\mu i}$, 
\bea
 e^{\epsilon \Lambda^{\mu i} \mathcal{L}_{{M}_{\mu i}} } e^{\phi^i(x) \mathcal{L}_{P_i}}g_{MN} = & \exp\left\{ (\phi^i + \epsilon \Lambda^{\mu i} x_\mu + \epsilon y^j \Lambda^{\mu j} \partial_\mu\phi^i) \mathcal{L}_{P_i} \right. \cr &\phantom{\exp \, \ }  \left.+ \epsilon \phi^i \mathcal{L}_{\com{{\Lambda}}{P_i}} + \mathcal{O}(\epsilon^2) \right\} g_{MN},
\eea
where we have defined ${\Lambda} = \Lambda^{\mu i} {M}_{\mu i}$. In writing this expression, we are omitting the compactly supported diffeomorphisms needed to to generate normalizable zero modes. This is just to avoid clutter. We are also using the inverse Higgs constraint from section~\ref{inversehiggs}\ to express the action of $\mathcal{L}_{M_{\mu i}}$ in terms of $\cL_{P_i}$ up to a non-normalizable bulk fluctuation. The bulk fluctuation is orthogonal to the normalizable zero modes and what we want to extract is the action of the symmetry on the normalizable `brane' modes.

The commutator is simple to evaluate, 
\be
\com{{\Lambda}}{P_i} = \Lambda^{\mu}{}_i \left({\partial \over \partial x^\mu} \right).
\ee
This is a manifest isometry of $g_{MN}$. An application of the Baker-Campbell-Hausdorff formula allows us to pull this to the right of the main exponential, where it acts trivially on $g_{MN}$, yielding:
\bea
e^{\epsilon \Lambda^{\mu i} \mathcal{L}_{M_{\mu i}} } e^{\phi^i(x) \mathcal{L}_{P_i}}g_{MN} = & \exp\left\{ (\phi^i + \epsilon \Lambda^{\mu i} x_\mu + \epsilon y^j \Lambda^{\mu j} \partial_\mu\phi^i \right. \cr & \left. - {1\over 2} \epsilon \phi^j \Lambda^{\mu j} \partial_\mu \phi^i ) \mathcal{L}_{P_i} + \mathcal{O}(\epsilon^2) \right\} g_{MN}.
\eea
We interpret the object in parentheses above as the transformation of $\phi^i$ under a boost. However, as before, the action of a boost includes a perturbation that is not normalizable in the asymptotic region of large $y$. Specifically, the metric perturbation generated by the operator $y^j \Lambda^{\mu j} \partial_\mu\phi^i \mathcal{L}_{P_i}$ decays like ${1\over y}$ as $y \to \infty$. 

Another way to note that this perturbation is orthogonal to the normalizable zero modes 
$h^{(i)}_{mn}$ with respect to the inner product $\langle \cdot , \cdot \rangle$ given in~\C{innerprod}\ is because of parity; the zero modes $h^{(i)}$ are odd under ${\vec y} \to -{\vec y}$, while the perturbation generated by $y^j \Lambda^{\mu j} \partial_\mu\phi^i \mathcal{L}_{P_i}$ is even.
Therefore, acting on the zero mode, boosts induce precisely the transformation found in the brane analysis of section~\ref{branescalars}, and needed to give the Dirac action, up to additional non-normalizable bulk metric perturbations:
\be
\phi^i \to \phi^i + \epsilon \Lambda^{\mu i} x_\mu  - {1\over 2} \epsilon \phi^j \Lambda^{\mu j} \partial_\mu \phi^i + \mathcal{O}(\epsilon^2).
\ee

\subsubsection{Including a gauge field}\label{includinggravgauge}

Our discussion so far has been concerned with scalar degrees of freedom. In section~\ref{includinggaugefields}, we described how the action for gauge-fields on branes is also constrained by the broken global symmetries. In this section, we turn our attention to gauge-fields in the gravity context and, specifically, for our example of the Taub-NUT space. 

Let us start by describing how gauge-fields emerge from Kaluza-Klein reduction of $11$-dimensional supergravity on a Taub-NUT space. Similar comments apply to type IIA string theory. For a nice discussion of this Kaluza-Klein reduction and the leading terms in the effective action for the gauge-fields, see~\cite{Imamura:1997ss}. The $11$-dimensional supergravity theory contains a metric $g$ and the $3$-form potential $C_3$ as bosonic fields. The bosonic terms in the supergravity action take the form
\be
S = \frac{1}{2\kappa_{11}^2} \int d^{11}x \sqrt{-g} \left(  \cR - \frac{1}{2} |G_4|^2\right)  - \frac{1}{2\kappa_{11}^2} \int \frac{1}{6} C_3 \w G_4 \w G_4 ,\label{eqn:action}
\ee
where $\cR_{MN}$ is the Ricci tensor and $\cR = g^{MN} \cR_{MN}$ is the Ricci scalar. 
The equations of motion that follow from~\C{eqn:action} are:
\bea
\cR_{MN}    &= \frac{1}{12} \left( G_{MPQR} G_{N}^{~~PQR} -2\,g_{MN} |G_4|^2\right).\label{eqn:einstein}\\
d\star G_4 &= -\half G_4 \w G_4,\label{f4eom}
\eea
where the norm of a rank $p$ tensor is defined by
\be\non
|T|^2 = {1\over p!} g^{M_1N_1} \cdots g^{M_p N_p} T_{M_1\ldots M_p} T_{N_1 \ldots N_p}.
\ee In the absence of explicit M5-brane sources, the field strength also satisfies the Bianchi identity $dG_4 = 0$. Our ansatz~\C{fourformansatz}\ for the potential gives a field strength $G_4=dC_3$ which automatically satisfies this constraint. 

 In a pure metric background like the Taub-NUT solution, the Chern-Simons coupling of~\C{eqn:action}\ is irrelevant and localized zero-modes for $C_3$ correspond to normalizable harmonic forms. Fortunately there is a unique normalizable $2$-form. Let $\omega_2$ denote this closed $2$-form.  Uniqueness implies that it must be either self-dual or anti-self-dual on the $4$-dimensional Taub-NUT space, and indeed it is anti-self-dual with our metric convention 
\be
\o_2 = -\ast \o_2. 
\ee
On a non-compact space like Taub-NUT, this form can always be trivialized:
\be
\o_2 = d\xi_1. 
\ee
The $1$-form $\xi_1$ is nicely expressed in terms of the variables~\C{refinedansatz},\footnote{We are using a slightly different choice of normalization for $\xi_1$ from the choice found in~\cite{Dasgupta:2003us}. The difference is only a numerical factor. } 
\be\label{definexi}
\xi_1 = {1\over H} \left( d\chi + \cos\th d\psi\right).
\ee
We can decompose $C_3$ into a $7$-dimensional gauge-field together with a closed $2$-form $B_2$ as follows, 
\be
C_3 = A_1(x) \wedge \o_2 + B_2(x) \wedge \xi_1. 
\ee
The gauge symmetry for $C_3$ consists of shifts by an exact $2$-form
\be
C_3 \rightarrow C_3 + d\La_2.
\ee
Imagine space-time-dependent gauge transformations of the form, 
\be\label{gauge}
\La_2 = \l_0(x) \o_2 + \l_1(x) \xi_1, 
\ee
where $\l_p$  denotes an $p$-form, rather than more general $(x,y)$-dependent gauge transformations. 
Under such transformations, 
\be
A_1 \rightarrow A_1 + d\l_0, \qquad B_2\rightarrow B_2 + d\l_1, \quad A_1 \rightarrow A_1 - \l_1.
\ee
Defining the field strength $F_2=dA_1$, we see that only the combination $F_2 + B_2$ is invariant under $\l_1$ gauge transformations. This is a familiar structure from the physics of D-branes. 

For the most part, we will not be concerned with $\l_1$ gauge transformations, but rather our interest will focus on $A_1$ and the standard abelian gauge transformations parametrized by $\l_0$. We need to determine how the large diffeomorphisms that generate the broken global symmetries act on $C_3$, and therefore on $A_1$. 

First we would like to observe that $A_1$ can be viewed as a Nambu-Goldstone vector for a broken global $2$-form symmetry. In this respect, it is quite similar to the universal NG bosons we discussed earlier. The following comments are motivated by a similar viewpoint expressed in~\cite{Adawi:1998ta}. For our pure metric background configuration, we chose the potential $C_3=0$. The choice that $C_3$ vanish identically is not gauge-invariant. We can choose  any other potential $C_3 = d \Xi_2$, as long as $\Xi_2$ vanish at infinity, i.e., that this is a true gauge transformation. Choosing $C_3 = 0$ asymptotically, however, spontaneously breaks the corresponding global symmetry under which $C_3$ transforms with a parameter $\Xi_2$ that does not vanish asymptotically.

In this background, there exists such a global transformation given by
\be\label{definelargeX}
\Xi_2 = A_1 \wedge \xi_1,
\ee
where $A_1$ is now taken to be a constant $1$-form transverse to the Taub-NUT metric and $\xi_1$ is defined in~\C{definexi}. The key point is that $d \xi_1 = \omega_2$ is normalizable, although $\xi_1$ is not normalizable. 
Using~\C{definelargeX}, the transformation of $C_3$ is
\be
\delta C_3 = A_1 \wedge \omega_2.
\ee
This is not a gauge transformation because $\xi_1$ does not vanish at infinity; however, because $\omega_2$ is closed, the gauge-invariant field strength $G_4$ is unchanged: $\delta G_4 = 0$. This in analogy to the scalar case considered previously. In that case, for a constant $\phi$, the metric
\be
e^{\phi \mathcal{L}_{P_y}} g_{MN}
\ee
is not related to the original metric by a gauge transformation, because the vector field $P_y$ does not vanish at infinity. However, this metric trivially still satisfies the equation of motion, because it only differs from the original metric by a Lie derivative. The next step, in this scalar case, was to enhance $\phi$ to a function of the Minkowski coordinates $\phi(x)$. The analogous promotion makes the constant $A_1$ a space-time field. From this perspective, $A_1$ and the NG-bosons are on similar footing. Alternately, one can simply view $A_1$ as arising from Kaluza-Klein reduction.

Now consider the perturbation, 
\be
\delta C_3 = A_1(x) \wedge \omega_2 \quad  \Rightarrow \quad \delta G_4 = F_2 \wedge \omega_2.
\ee
Plugging into the action and integrating over the Taub-NUT space gives the  equation of motion for $F_2$:
\be
d \ast_7 F_2 = 0,
\ee
where $\ast_7$ is the Hodge operator in the $7$-dimensional Minkowski space. This equation of motion is only true to leading order in the fluctuation. The full equations of motion~\C{eqn:einstein}\ and~\C{f4eom}\ are highly non-linear. 

How does $A_1$ transform under the broken Lorentz symmetry? For the scalar field, we found the correct transformation by simply acting with $\mathcal{L}_{\Lambda}$ on the metric. Let us try the same approach here. We will use two nice identities for the action of Lie derivatives: ``Cartan's magic formula" and a simple corollary:
\be
\bes
\mathcal{L}_{\zeta} \eta_p &= \acom{d}{\iota_\zeta} \eta_p, \cr
\mathcal{L}_{\zeta}\left( \eta_p \wedge \omega_q \right) &= \left( \mathcal{L}_\zeta \eta_p \right) \wedge \omega_q + \eta_p \wedge \mathcal{L}_\zeta \omega_q.
\ees
\ee
In these expressions, $\zeta$ is some vector field while $\eta_p$ and $\omega_q$ are form fields of denoted degree. Moving onto business, we compute:
\be
\mathcal{L}_\Lambda \left( A_1 \wedge \omega_2 \right) = \iota_\Lambda F_2 \wedge \omega_2 + F_2 \wedge \iota_\Lambda \omega_2.
\ee
Comparing with~\C{branegaugetransformation}, we see that neither of these terms should be there since we have yet to wiggle the metric so effectively  $\phi = 0$ in this setup. Can we explain them away? The first term,
\be
\left( \iota_\Lambda F_2\right)_\mu \sim y^i \Lambda^{\nu i} F_{\nu \mu},
\ee
has explicit $y$-dependence. The corresponding form $y^i \o_2$ is therefore orthogonal to the harmonic $2$-form in much the same way as we saw for the scalar. In other words, this is a transformation of either massive or non-normalizable modes, but not of $A_1$. The reasoning for the second term is similar. Specifically, $\iota_\Lambda \omega_2$ is a $1$-form on Taub-NUT which we can write in the form
\be\label{secondtermiota}
\iota_\Lambda \omega_2 = \cL_{\Lambda} \xi_1 - d\left( \iota_\Lambda \xi_1\right). 
\ee
The first term on the right hand side of~\C{secondtermiota}\ is non-normalizable, and the second term is a closed $1$-form. However, there are no normalizable closed $1$-forms on Taub-NUT. Again we find a transformation of $A_1$ into bulk modes under the boost.



What about for non-zero $\phi(x)$? In this case, we introduce a wiggled version of the metric, as before, and a $\phi$-dependent version of $\delta C_3$:
\be
\{ g_{MN}(\phi),~ \delta C_3(\phi) \} = \{e^{\phi^i(x) \mathcal{L}_{P_i}} g_{MN},~ A_1\wedge \omega_2(\phi) \}.
\ee
Here we have defined the wiggled $2$-form as follows:
\be
\omega_2 (\phi) = d \left( e^{\phi^i(x) \mathcal{L}_{P_i}} \xi_1 \right).
\ee
The advantage of this definition is that $\o_2(\phi)$ is still a closed form. Now we want to see how this configuration is modified by the boost operator,
\be
e^{\epsilon \mathcal{L}_{\Lambda}} \left\{ e^{\phi^i(x) \mathcal{L}_{P_i}} g_{MN},~ \, A_1\wedge  \omega_2(\phi)\right\} = \left\{e^{ \phi'^i(x) \mathcal{L}_{P_i}} g_{MN},~ \, A_1'\wedge \omega_2(\phi') \right\},
\ee
where $\phi'$ and $A_1'$ are the transformed fields. In particular, we found previously that when $A_1 = 0$, 
\be
\phi'^i = \phi^i + \epsilon \nabla_{\Lambda} \phi^i + \epsilon (\Lambda \cdot x)^i  - {1 \over 2} \epsilon \nabla_{\Lambda \cdot \phi} \phi^i  + \mathcal{O}(\epsilon^2).
\ee
So we want to analyze
\be\label{boostedform}
e^{\epsilon \mathcal{L}_{\Lambda} }\left(A_1\wedge \omega_2(\phi) \right).
\ee
This splits into two terms,
\be\label{twopieces}
\left( e^{\epsilon \mathcal{L}_\Lambda} A_1 \right) \wedge \omega_2(\phi) + A_1 \wedge d \left( e^{\epsilon \mathcal{L}_\Lambda} e^{\phi^i(x) \mathcal{L}_{P_i}} \xi_1 \right),
\ee
where in the second term we have used the observation that $d$ commutes with the Lie derivative. Furthermore, in the second term we see the combination of exponentials that enacted the change from $\phi^i$ to $\phi'^i$ in the metric. Again, the BCH relation combined with the inverse Higgs constraint enacts this change in this case. To be more explicit, at linear order in $\epsilon$, the modified form field in~\C{boostedform}\ can be expressed as
\be
\left(\delta A_1\right) \wedge \omega_2(\phi) + A_1 \wedge \omega_2(\delta \phi),
\ee
where both $\delta A_1$ and $\delta \phi$ are linear in $\epsilon$. So, to examine the transformation of $A_1$, namely $\delta A_1$ we are interested in the first term above, which by the previous argument corresponds to to the first term in~\C{twopieces}.

Expanding the Lie derivative, and working only to order $\epsilon$, the first term in~\C{twopieces}\ is given by 
\be
\left( \iota_\Lambda F_2 \wedge +  d( \iota_\Lambda A_1) \right) \wedge d \left( e^{\phi^i(x) \mathcal{L}_{P_i}} \xi_1 \right).
\ee
At this point, we also want to work only to first order in $\phi$ since the transformations with which we are comparing~\C{branegaugetransformation}\ are only linear in $\phi$. So we expand the exponential and ignore the $\phi$-independent term, because this will be the same as in the case when $\phi = 0$, explained previously. We find, 
\be
\iota_\Lambda F_2 \wedge d\left( \phi^i(x) \mathcal{L}_{P_i} \xi_1 \right) + d( \iota_\Lambda A_1) \wedge d \left( \phi^i(x) \mathcal{L}_{P_i} \xi_1 \right).
\ee
The second term is exact and corresponds to a gauge transformation of $C_3$ so we can drop it. This leaves
\be
\iota_\Lambda F_2 \wedge d\phi^i(x) \wedge \mathcal{L}_{P_i} \xi_1 + \phi^i(x) \iota_\Lambda F_2 \wedge \mathcal{L}_{P_i} \omega_2.
\ee
Finally, we can `integrate by parts' to move the $\mathcal{L}_{P_i}$ operator in the following way. In addition to being able to ignore exact terms, we claim that we can also ignore terms that are $\mathcal{L}_{P_i}$ of anything. Since the only $y$-dependence comes from $\omega_2$ or $\xi_1$ and the interior product $\iota_\Lambda$ acting on a space-time form,  integration by parts has the effect of transferring the $\mathcal{L}_{P_i}$ operator from $\xi_1$ to $\iota_\Lambda F_2$. 
This leaves us with
\be
\mathcal{L}_{P_i}\left( \iota_\Lambda F_2\right) \wedge d\phi^i(x) \wedge  \xi_1 + \phi^i(x)  \mathcal{L}_{P_i} \left( \iota_\Lambda F_2 \right) \wedge \omega_2.
\ee
At this point, we note that the first term is a space-time form wedged with $\xi_1$. It represents a non-normalizable fluctuation that we will neglect as we have been consistently doing for all other non-normalizable modes.

The second term is finally of the form $\delta A_1 \wedge \omega_2$, and so we can extract
\be
\delta A_1 = \phi^i(x) \mathcal{L}_{P_i} \left( \iota_\Lambda F_2 \right).
\ee
We can compare to the transformation in~\C{branegaugetransformation} by writing this in components:
\be
\delta A_\mu = \phi^i \Lambda^{\nu i} F_{\nu \mu}. 
\ee
This agrees with~\C{branegaugetransformation}\ up to a field-dependent $A_1$ gauge transformation. From gravity, we at least recover the constraints seen from a brane analysis. In the following section, we will turn to the question of whether we can see more than the constraints found in a brane analysis.

\section{Seeing A Critical Electric Field}\label{critelectric}

We would now like to examine whether gravity gives us more data about the equation of motion for the gauge-field. The broken Poincar\'e transformations reproduce the same data found in a brane analysis, but those constraints are insufficient to imply the Born-Infeld action for the abelian gauge-field $A_1$. A striking difference between the Born-Infeld theory and a free Maxwell theory is the existence of a critical electric field. Any constant electric field is a solution of the Maxwell theory, but there is a maximum value permitted by~\C{DBI}. If we can see a critical value emerge from gravity, we will have strong evidence that standard two derivative gravity on a Taub-NUT space gives rise to the Born-Infeld theory.  This has to be quite non-trivial to see because a critical value for the electric field $E$ cannot be seen in a perturbative expansion of 
\be
{1\over \sqrt{1-E^2}} 
\ee
around small values of $E$. Rather, we need data about the back-reacted gravitational solution for finite $E$. 

Imagine turning on a constant electric field $E$ for $A_1$ in the $x_1$ direction. The presence of this field strength will deform the Taub-NUT solution. Kaluza-Klein reduction on Taub-NUT does not give a localized graviton. Gravity is a bulk phenomena; the brane modes only consist of $A_1$ together with the NG-bosons. The back-reaction from $E$ should deform only the Taub-NUT directions, and possibly distinguish the $(x_0, x_1)$ directions relative to the other spatial $x$ directions. Symmetry requires the deformation to depend only on the radial coordinate $y$. With these considerations, the deformed metric takes the generic form   
\bea \label{metricansatz}
ds^2_{\rm electric} =& \, H_1(y) \left\{ - dx_0^2 + dx_1^2 \right\} + H_2(y) \left\{ dx_2^2 +\ldots+ dx_6^2 \right\} \cr & + H_3(y) \left(  dy^2 + y^2 d\th^2 + y^2 \sin\th^2 d\psi^2\right) + H_4(y)^{-1} \left( d\chi + \cos\th d\psi \right)^2. 
\eea
In the prior case without an electric field, 
\be
H_1= H_2=1, \qquad H_3=H_4 = H(y) = 1 + {1\over y}, 
\ee
where we have chosen $\vec{y}_0=0$. As in the discussion of section~\ref{includinggravgauge}, we define 
\be
\xi_1 = d\chi + \cos\th d\psi.
\ee
An electric field with magnitude proportional to a constant $E$ corresponds to a background potential
\be\label{fourformansatz}
C_3 = E \, H_5(y) \, dx^0\wedge dx^1\wedge \xi_1, 
\ee
where $H_5(y) \rightarrow 1$ as $y\rightarrow \infty$. It is this background potential~\C{fourformansatz}\ which produces the warping of the metric~\C{metricansatz}. 

A direct analysis of the Einstein equations involves solving for the $5$ functions, $H_i(y)$, of the single variable $y$.  This is not an easy task, but not a completely hopeless task because of a lot of data can be gleaned by studying the leading ${1\over y}$ terms in the expansion of each $H_i$. However, we can use string theory as a solution generating technique to arrive at the fully back-reacted solution. Specifically, T-duality can be used to generate constant magnetic or electric fields from toroidal metric configuration with off-diagonal metric terms. This approach was used in~\cite{Dasgupta:2003us}\ to generate deformed Taub-NUT solutions with magnetic fields, following the earlier work of~\cite{Maldacena:1999mh, Hashimoto:1999ut} \ in the context of other branes with background fluxes.

With this technique, the five $H_i(y)$ can be written in terms of two functions $h_1(y)$ and $h_2(y)$ depending only on a single parameter which we call $\alpha$:
\be\label{functionassignment}
H_1 = h_1^{2\over 3} h_2^{-{2\over 3}}, \, \ H_2 = h_1^{-{1\over 3}} h_2^{{1\over 3}}, \, \ H_3 = h_1^{2\over 3} h_2^{1\over 3}, \, \ H_4 = h_1^{1\over 3} h_2^{2\over 3}, \, \ H_5 = h_2^{-1}.
\ee
And
\be
h_1(y) = 1 + {\alpha \over y}, \quad h_2(y)  = 1 + {1 \over \alpha y}.
\ee
Note that if $\alpha = 1$ and $E = 0$, we reproduce the undeformed Taub-NUT solution. We would like to see how the equations of motion determine $\alpha$ in terms of $E$ when the latter is non-zero.

Let us first check that the ansatz~\C{fourformansatz}\ combined with the assignment~\C{functionassignment}\ solves the tensor field equation of motion. The $G_4\wedge G_4$ term in the equation of motion vanishes so we need only focus on the left hand side of~\C{f4eom}.  To compute the left hand side, take a convenient orthonormal basis
\bea
&e_0 = {h_1(y)^{{1\over 3}} h_2(y)^{-{1\over 3}} dx^0}, \qquad e_1 = {h_1(y)^{{1\over 3}} h_2(y)^{-{1\over 3}} dx^1}, \cr & e_\mu = {h_1(y)^{- {1\over 6}} h_2(y)^{{1\over 6}} dx^\mu}, \qquad \mu=2,\ldots 6, \cr
 & e_y = {h_1(y)^{{1\over 3}} h_2(y)^{{1\over 6}} dy}, \quad e_{\theta} = h_1(y)^{{1\over 3}} h_2(y)^{{1\over 6}} y d\theta,  \quad e_{\psi} =h_1(y)^{{1\over 3}} h_2(y)^{{1\over 6}} y \sin\th d\psi, \cr
 & e_{\chi} = h_1(y)^{-{1\over 6}} h_2(y)^{-{1\over 3}} \left( d\chi + \cos\theta d\psi \right). \eea
The metric expressed in this basis is $ds^2_{\rm electric} = e_\mu e^\mu + e_y^2 + {e_\th}^2 + {e_\psi}^2 + {e_\chi}^2$ and we determine an orientation by choosing a volume form: 
\be
e_0\wedge  \ldots \wedge e_6 \wedge e_y \wedge e_\th \wedge e_\psi \wedge e_\chi. 
\ee
The field strength $G_4$ takes the form
\be\label{explicitG}
G_4 = - {E} e_0\wedge e_1 \wedge \left\{  {\left({h_2' \over h_1^{5\over 6} h_2^{7\over 6}} \right)} e_y e_\chi  + \left( {1 \over  y^2 h_1^{4\over 3} h_2^{2\over 3}} \right) e_\th e_\psi \right\}.
\ee 
Taking the Hodge dual gives,
\be
\star G_4 = E h_1^{-{5\over 6}} h_2^{5\over 6} \, dx^2 \wedge \ldots dx^6 \wedge \left\{  {\left({h_2' \over h_1^{5\over 6} h_2^{7\over 6}} \right)}  e_\th e_\psi + \left( {1 \over  y^2 h_1^{4\over 3} h_2^{2\over 3}} \right)  e_y e_\chi \right\}.
\ee
As a basic check, take $\alpha=1$ which corresponds to the case of a vanilla Taub-NUT. For this choice, $h_1=h_2$ and the form is anti-self-dual in the Taub-NUT directions if
\be
h_2' = -{1\over y^2}, 
\ee
which is true. For general $\a$, the internal part of $G_4$ is no longer ant-self-dual, and we must instead check that $\ast G_4$ is closed. With the aid of a computer algebra program, we find that it is closed for any value of $\alpha$ and $E$. So, we must look to the Einstein equation to constrain $\alpha$ in terms $E$.

For simplicity, we will focus on the $2-2$ component of the Einstein equation~\C{eqn:einstein}\, which reads
\be
\mathcal{R}_{22} + {1 \over 6} g_{22} \left| G_4 \right|^2 = 0.
\ee
This component of the equation is especially simple because $G_{2 MNP}G_{2}{}^{MNP} = 0$. Again with the aid of computer algebra, we find that the equation is satisfied if
\be
-{\left(2 \alpha +\alpha ^2 y+y\right) \over 6 (\alpha +y)^3 (\alpha  y+1)^2}\left(E^2 - 1 + \alpha^2 \right)  = 0,
\ee
which, because $\alpha \geq 0$, is equivalent to
\be
E^2 = 1 - \alpha^2.
\ee
From this equation, we find the critical value of $E$, which is $1$ in our units. Furthermore, this equation is satisfied is by
\be
\alpha = \textrm{sech} ~\!\beta, \quad E = \textrm{tanh} ~\!\beta,
\ee
for some $\beta$.
This computation can be extended to include more general combinations of background magnetic and electric fields. It essentially implies the Born-Infeld action for the gauge-field, and when combined with our prior results on the NG-bosons, gives the DBI action~\C{DBI}\ directly from a gravitational analysis.


\section*{Acknowledgements}

It is our pleasure to thank Ferdinando Gliozzi, Cindy Keeler and David Kutasov for useful discussions. We would particularly like to thank Austin Joyce for early important participation in this project. S.~S. would like to thank the organizers of the ``String Theory in London'' 2016 meeting, and the organizers of the special session of the sectional meeting of the AMS at NCSU on ``Mathematical String Theory,'' where preliminary versions of this work were presented. 
 T.~M. and S.~S. are supported, in part, by NSF Grant No.~PHY-1316960. 
 
\newpage

\newpage

\ifx\undefined\bysame
\newcommand{\bysame}{\leavevmode\hbox to3em{\hrulefill}\,}
\fi

\end{document}